 \documentclass[final,times,onecolumn]{elsarticle}

\pdfoutput=1 

\usepackage{amssymb}
\usepackage{lipsum}
\usepackage{graphicx}
\usepackage[table,xcdraw]{xcolor}
\usepackage{subcaption}
\usepackage{enumitem}
\usepackage{graphicx}
\usepackage{booktabs}
\usepackage{float}
\usepackage{mathtools}
\usepackage{hyperref}
\usepackage{amsmath}
\usepackage{multicol}
\usepackage{multirow}
\usepackage{longtable}
\usepackage{tabu}
\usepackage[modulo]{lineno}
\usepackage{natbib}
\modulolinenumbers[5]
\usepackage{setspace}
\journal{Nuclear Physics A}

\begin{document}

\begin{frontmatter}

\title{Measurement of the Photon Detection Efficiency of Hamamatsu VUV4 SiPMs at Cryogenic Temperature}

\author{R.~\'{A}lvarez-Garrote$^a$}
\author{E.~Calvo$^a$}
\author{A.~Canto$^a$}
\author{J.~I.~Crespo-Anad\'{o}n$^a$}
\author{C.~Cuesta$^a$}
\author{A.~de~la~Torre~Rojo$^a$}
\author{I.~Gil-Botella$^a$}
\author{S.~Manthey~Corchado$^a$}
\author{I.~Mart\'{i}n$^a$}
\author{C.~Palomares$^a$}
\author{L.~P\'{e}rez-Molina$^a$}
\ead{laura.perez@ciemat.es}
\author{A.~Verdugo~de~Osa$^a$}

\address[first]{CIEMAT, 
           Avda. Complutense 40, 
           28040 
           Madrid 
           (Spain)}


\begin{abstract}
Liquid argon time projection chambers (TPC) are widely used in neutrino oscillation and dark matter experiments. Detection of scintillation light in liquid argon TPC's is challenging because of its short wavelength, in the VUV range, and the cryogenic temperatures ($\sim$86 K) at which the sensors must operate. Wavelength shifters (WLS) are typically needed to take advantage of the high Photon Detection Efficiency (PDE) in the visible range of most of photondetectors. 
The Hamamatsu VUV4 S13370--6075CN SiPMs can directly detect VUV light without the use of WLS, which main benefit is an improved PDE at these short wavelengths, but also the visible light from WLS.
The manufacturer (Hamamatsu Photonics K.K.) provides a complete characterization of these devices at room temperature; however, previous studies have indicated a decrease of the PDE at cryogenic temperature for VUV light. 
In this work, we present the measurement of the PDE of VUV4 SiPMs at cryogenic temperature for different wavelengths in the range [270, 570] nm. A dedicated measurement at 127 nm is also shown.
\end{abstract}

\end{frontmatter}

\section{Introduction} \label{sec:intro}

To optimize the energy resolution and sensitivity of the next generation of liquid argon time proyection chamber~(LArTPC) neutrino and direct dark matter detection experiments it is essential to maximize the light collection process~\cite{Rubbia_LArTPC_NewConcept(CERN-EP-INT-77-8),2108.01902_LArTPC_DUNE,1910.08218_LArTPCTriggerSBND,1707.08145_LArTPCDarkSide}. The main challenge relies on the detection of the LAr scintillation photons whose peak emission wavelength is centered at 127 nm~\cite{Heindl:2010}. These photons can be directly detected by VUV photodetectors or indirectly after using a wavelength--shifter~(WLS) typically around 420~nm. The cryogenic environment is also a constraint on the choice of the photosensor.

As the large scale of the new generation detectors requires the use of large area devices for light detection, these sensors can also be integrated in light collection devices~\cite{1805.00382_X-ARAPUCATest} or forming arrays~\cite{DarkSide-20k:2017} where WLS materials are commonly used.
Besides, the use of WLS dopants in LAr, like Xe, or the installation of WLS coated reflective foils in the TPC walls is becoming a general practice because of the dimensions of these giant detectors (several meters) and the relative short scattering length in LAr for light of 127~nm (1~m).
The inclusion of sensors sensitive to a large range of wavelengths is interesting for detecting shifted and direct light with the same device. The Hamamatsu VUV4 S13370 multi--pixel photon counting is a good candidate with a photon detection efficiency~(PDE) between 20\% and 40\% for wavelengths between 100 and 600~nm. 
In contrast to other SiPM models, for which it does not seem to vary~\cite{Iwai:2019, Biroth:2015}, previous studies have shown a decrease in the PDE of these SiPMs at cryogenic temperatures~($<$200~K) for 127~nm.  
In this paper, a  measurement of their PDE at cryogenic temperature~(CT) for wavelengths ranging from 127 to 570~nm is presented for the first time. The data could be of particular interest for experiments requiring multi--wavelength detection. Additionally, calibrated VUV4 SiPMs can be used in the measurement of the conversion efficiency of wavelength shifters and for the calibration of complex devices including wavelength shifting~\cite{Palomares:2022}.

The paper is organized as follows. In section~\ref{sec:vuv4_sipms} we introduce the VUV4 SiPMs characterized in this study. Then, in section~\ref{sec:methodology} the employed methodology is described, followed by the designed experimental setups in section~\ref{sec:setup}. Finally, in section~\ref{sec:results} the results of the VUV4 SiPMs measurements are presented, followed by concluding remarks in section~\ref{sec:conclusions}.

\section{Hamamatsu VUV4 SiPMs} \label{sec:vuv4_sipms}
\begin{figure}[H]
    \begin{subfigure}{0.45\linewidth}
        \centering
        \includegraphics[width=0.55\linewidth]{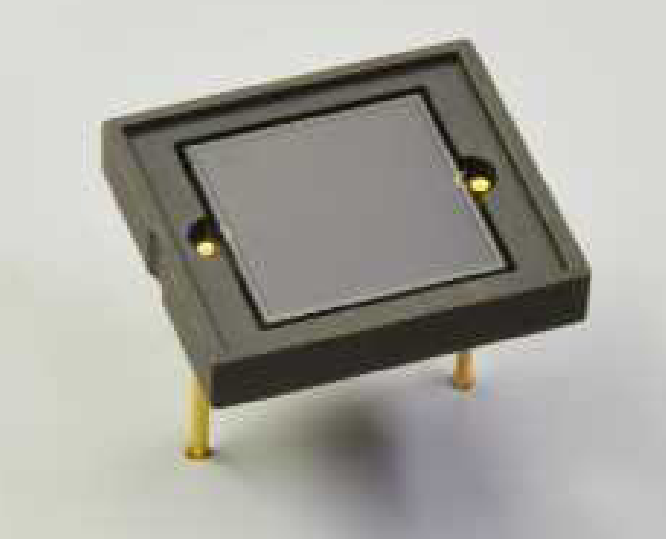}
        \caption{}
        \label{fig:sipm_photo}
    \end{subfigure}%
    \begin{subfigure}{0.45\linewidth}
        \centering
        \includegraphics[width=0.9\linewidth]{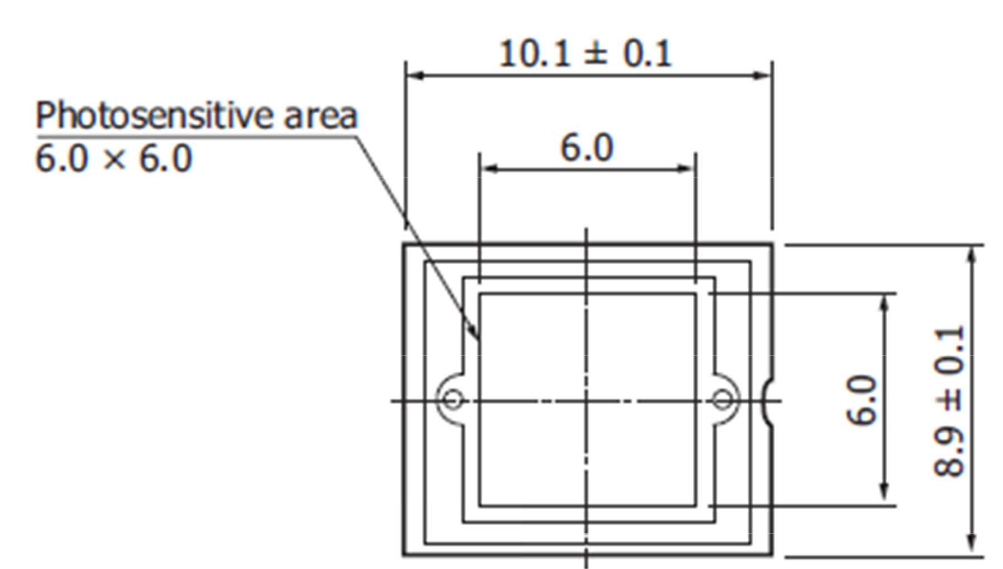}
        \caption{}
        \label{fig:sipm_scheme}
    \end{subfigure}
\caption{(a) Image of the S13370--6075CN SiPM model tested in this work. (b) Dimensional outline [mm]}
\label{fig:InfoSipm}
\end{figure}

The Hamamatsu (HPK) VUV4 SiPMs (series: S13370) are directly sensitive to the LAr scintillation light of 127~nm, 
and, moreover, they are prepared to carry out a stable performance at CT. Two VUV4 S13370--6075CN SiPMs have been studied in this work. In Figure~\ref{fig:InfoSipm} we can see a picture of one SiPM and its dimensional outline. 

These SiPMs have a micro--cell pitch of 75~$\rm\mu$m, an effective photosensitive area of $(6 \times 6)$ mm$^2$ with $70\%$ fill factor, and a typical breakdown voltage at room temperature~(RT) of $\sim$53~V. The devices have a ceramic package and no window. They also contain a metal quenching resistor to maintain its pulse shape at low temperatures. The main properties  provided by the manufacturer are summarized in Table~\ref{tab:KeyParameters}.

\begin{table}[!ht]
\centering
\begin{tabular}{ll} 
 \hline
    Parameter                     & Value                   \\ \hline \hline
    Effective photosensitive area & $(6 \times 6)$ mm$^2$   \\
    Size                          & $(10 \times 9)$ mm$^2$  \\
    Pixel pitch                   & $75$ $\rm \mu$m         \\
    Terminal capacitance          & $1.28$~nF               \\
    Fill factor                   & $70\%$                  \\
    Crosstalk probability         & $5$$\%$                 \\
    Gain                          & $5.8 \cdot 10^6$        \\
    Dark counts (typ.)            & 0.11~MHz                \\
    Breakdown voltage             & $(55 \pm 0.2)$~V        \\
    Breakdown voltage [86.15 K]   & $(42 \pm 0.2)$~V        \\ \hline
    \end{tabular}
    \caption{Key parameters of the HPK VUV4 SiPMs (series: S13370) provided by Hamamatsu Photonics K.K. at an over--voltage of 4~V. RT is assumed unless otherwise specified.}
    \label{tab:KeyParameters}
\end{table}

\begin{figure}[H]
    \centering
    \includegraphics[width=\linewidth]{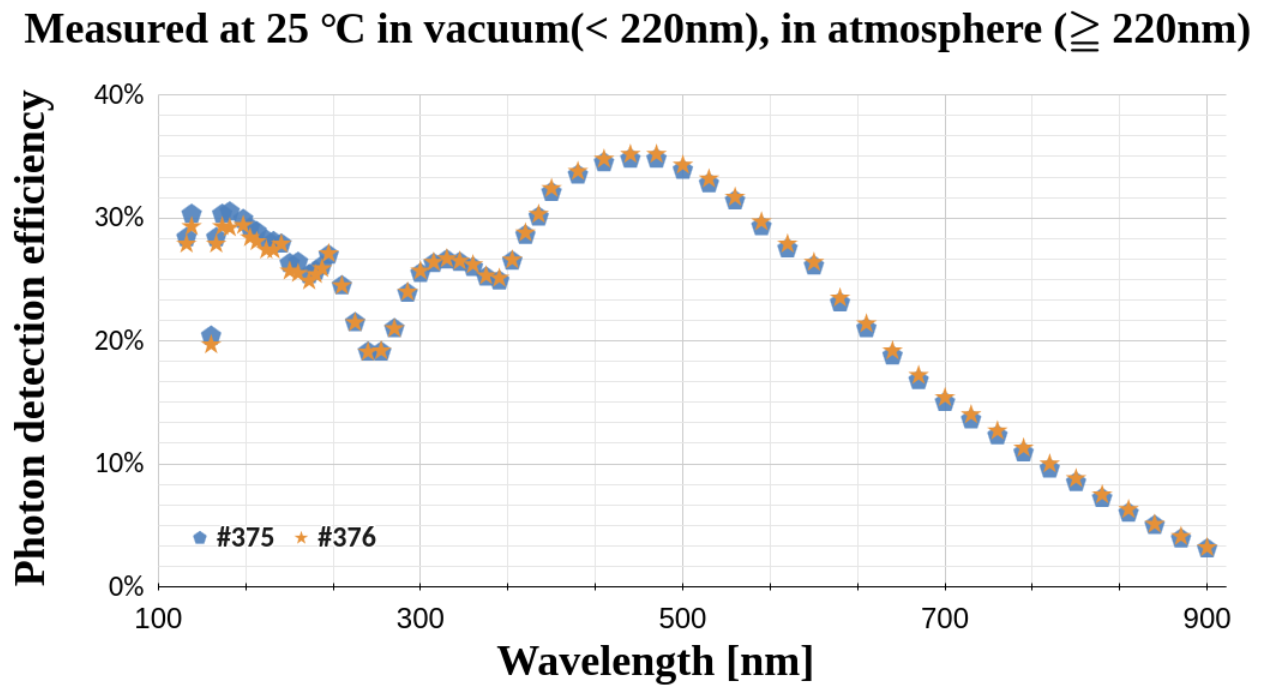}
    \caption{Photon detection efficiency at RT and OV = 4~V from 120~nm to 900~nm light measured and provided by Hamamatsu Photonics K.K.. The legend corresponds to the serial numbers of the SiPM model tested (blue pentagons \#375 and orange stars \#376).}
    \label{fig:HPK_PDEs_RT}
\end{figure}

The SiPMs have been calibrated by the manufacturer at RT at over--voltage (OV) of 4~V. The PDE for wavelengths from 120~nm to 900~nm measured by HPK is shown in Figure~\ref{fig:HPK_PDEs_RT}.
These SiPMs 
have been considered by several experiments involving liquefied noble gases~\cite{1903.03663_VUV4S13370nEXO,Parsa2022}
and, therefore, there are studies in the literature where key parameters and findings at CT are reported. Some of the them are of particular interest to our study and are summarized as follows:
\begin{itemize}
    \item At a temperature of 163~K and 3.1~OV, the dark noise was measured to be (0.137 $\pm$ 0.002)~Hz/mm$^2$. The correlated avalanche noise (i.e., crosstalk and afterpulses) is measured to be about 16\%, three times larger than at RT (see~\cite{1903.03663_VUV4S13370nEXO}).
    \item The reflectivity of these devices in liquid Xe has been measured in \cite{1910.06438_Reflectivity+PDE_HPK_VUV4SiPMs_LXe,2104.07997_ReflectivityVUVSiPMsS13370_LXe}. The specular reflectivity at 15$^{\circ}$ incidence of three samples of VUV4 SiPMs was found to be around 29\%. The same measurement in LAr gave similar results (see~\cite{1912.01841_ReflectanceVUVS13370-6075CN}). This parameter can modify the effective PDE if it is not subtracted in its measurement.
    \item Finally, at an OV of 3.91~V and a wavelength of 127~nm in liquid argon at 91.2~K, the PDE was measured as 14.7$^{+1.1}_{-2.4}$\% (see~\cite{2202.02977_SiPMVUV4_PDE_CT+128nm}). There are no measurements of the PDE at CT for wavelengths above the VUV range.
\end{itemize}
\section{Methodology} \label{sec:methodology}

\subsection{PDE measurement for visible light}
The PDE will be measured as the ratio of detected to incident photons. In this case, the photons reflected in the SiPM surface are not subtracted and, therefore, the measured PDE will depend on the incident angle.
The measurement of the PDE at CT for HPK VUV4 SiPM has been made using as reference the known PDE at RT provided by the manufacturer using equation~\ref{eq:PDE_CT}. The method is based on the comparison of the light detected by the SiPM at different temperatures when exposed to light of wavelengths in the range [270, 570]~nm. 
\begin{equation}
    \text{PDE}_{\text{CT}} = \frac{\text{Detected light (CT)}}{\text{Detected light (RT)}} \cdot \text{PDE}_{\text{RT}} \quad .
    \label{eq:PDE_CT}
\end{equation}
This methodology requires a controlled light source that emits a reproducible number of photons.
Due to the dependence of the SiPM gain with temperature, we perform calibration measurements at the single photoelectron~(SPE) level for each studied temperature (298~K and 77~K). The SiPM is able to distinguish signals from one, two or more photoelectrons (PE) and thus we can calculate the gain by acquiring a charge spectrum. These measurements are always performed at three different OV so the linear behavior of the gain can be verified. The PDE is calculated using a higher intensity light regime (between 50--150~PEs) and for an over-voltage of 4~V, which is the OV recommended by the manufacturer. 
The followed procedure was:
\begin{enumerate}
    \item Gain calibration measurements at RT for three different OV values.
    \item High-intensity light pulses at RT for different wavelengths between [270--570]~nm.
    \item Cool down of the system with liquid nitrogen (LN$_2$).
    \item Gain calibration measurements at CT for three different OV values.
    \item High-intensity light pulses at CT for different wavelengths between [270--570]~nm.
\end{enumerate}

\subsection{PDE measurement for 127~nm light}

The measurement at 127~nm requires the propagation of the light in vacuum from the source, making the measurement much more complex. An alternative method~\cite{2008.05371_ARAPUCA_efficiency,2104.07548_X-ARAPUCA_PD_DUNE} consists of the generation of 127~nm light directly in LAr with a radioactive source and comparing the detected number of PE with the theoretical expectation taking into account the ionizing particle, its energy, and the LAr scintillation light--yield.
This methodology requires the use of simulations for calculating the solid angle of the sensor with respect to the generated light. The effect of the LAr purity on the light--yield should also be taken into account.
\section{Experimental Setups} \label{sec:setup}

\subsection{Setups for PDE measurement for visible light}
Two different setups have been developed for this measurement, a tube at vacuum equipped with a heat exchanger 
that will be our reference configuration 
and another one, where the tube is filled with gas argon~(GAr), used as cross--check.

\begin{figure}[!ht]
    \centering
    \includegraphics[width=\linewidth]{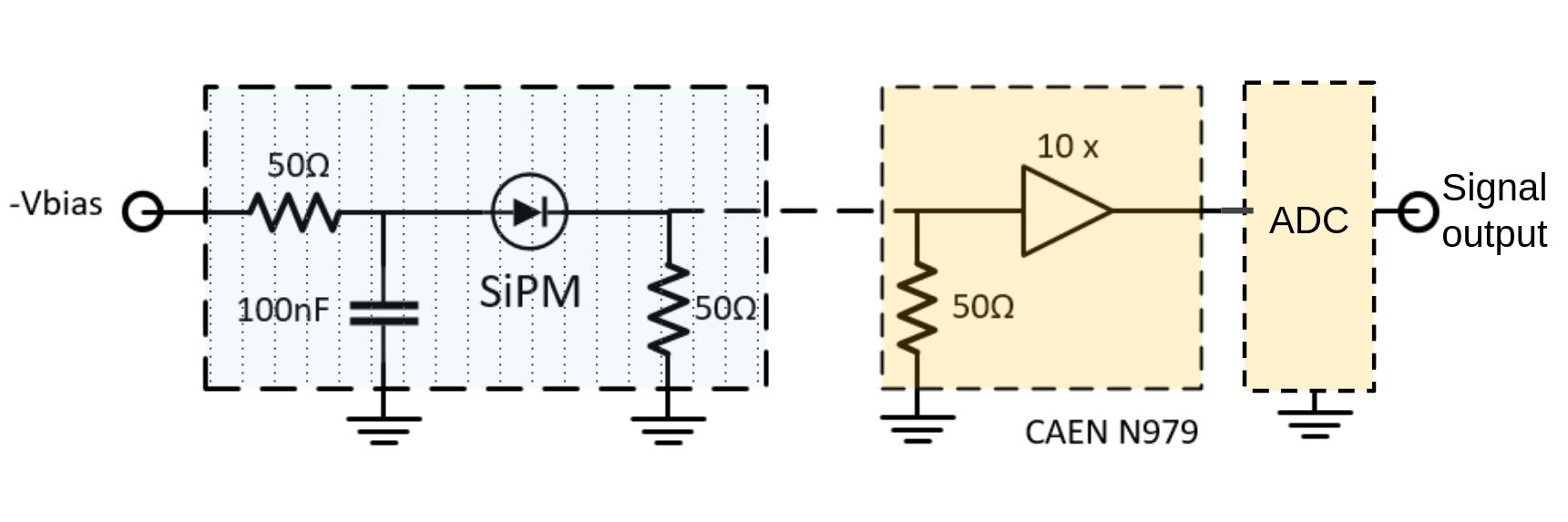}
    \caption{Diagram of the electronic circuit. The dotted blue shaded side is the front--end electronics, which will be at the same temperature than the SiPM (left side); while the orange, where the digitization is included, will be always at room temperature (right side).}
\label{fig:electronic_circuit}
\end{figure}

The read--out of the SiPM signal is the same in both cases and is shown in Figure~\ref{fig:electronic_circuit}.

The front--end electronics contains an amplifier, the voltage produced by the SiPM at the input of the amplifier is the result of two 50~$\Omega$ resistors connected in parallel. The voltage gain of the amplifier is 10 and the total trans-impedance gain is $V_{\text{output}}/I_{\text{SiPM}}=250\ \Omega$. In the right part of the diagram we can see that the ADC (model CAEN DT5725S~\cite{CAEN}) is connected to an amplifier (CAEN N979 model~\cite{CAEN}). The signal output is finally acquired with the ADC.

The trigger signal is provided by a pulse generator (model Aim-Tti 2511A) that is synchronously sent to the ADC and the light source. The ADC provides the digitized waveforms with 4~ns time and 14--bits resolution from the SiPM signal. 

\begin{figure}[!ht]
    \centering
    \includegraphics[width=0.7\linewidth]{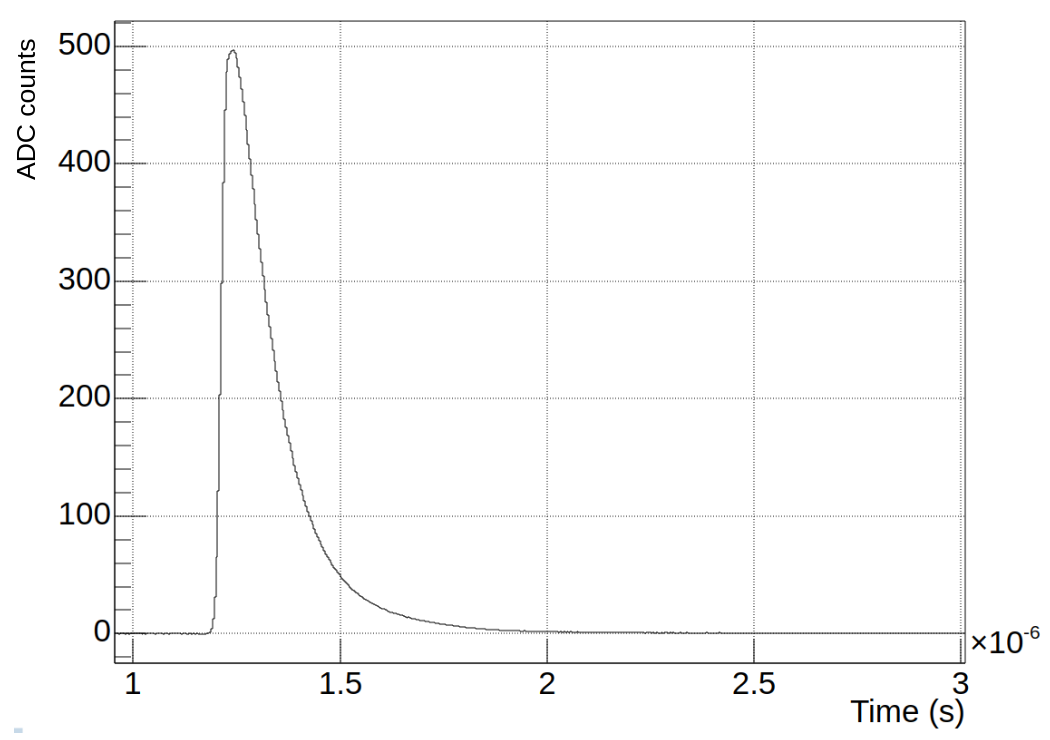}
    \caption{Average SiPM waveform obtained with the ADC at CT of 20000 pulses.}
    \label{fig:average_waveform}
\end{figure}

Figure~\ref{fig:average_waveform} shows an example average waveform of 20000~events corresponding to 270~nm LED light pulses obtained with the ADC at CT.

For measuring the PDE at different wavelengths we have made use of different LEDs and lasers to illuminate the SiPM. Table~\ref{tab:leds_info} shows the manufacturers and reference numbers, and the operational wavelengths of these light sources. 

\begin{table}[!ht]
\centering
    \begin{tabular}{llc} 
    \hline
    Type & Model & $\lambda$ (nm) \\ \hline \hline
    LED & Roithner -- UVR270-SA3P & 270 \\
    LED & Roithner -- UVR280-SA3P & 280 \\
    LED & ThorLabs -- LED315W & 317 \\
    LED & Roithner -- XSL-355-3E-R6 & 355 \\
    LED & Roithner -- LED385-33 & 385 \\
    Laser & PicoQuant -- PDL828+LDH-P-C-405 & 405 \\
    Laser & PicoQuant -- PDL828+LDH-P-C-420 & 420 \\
    LED & OSRAM -- LB 543C-VAW-35 & 470 \\
    LED & Vishay -- TLHG4200 & 570 \\ \hline
    \end{tabular}%
    \caption{Information about manufacturers, reference number and wavelength of the used LEDs and the setup where they are used. The references for the manufacturers can be found in~\cite{Thorlabs,Roithner,PicoQuant,OSRAM,Vishay}}
    \label{tab:leds_info}
\end{table}

\begin{figure}[htp]
    \centering
    \includegraphics[width=0.7\linewidth]{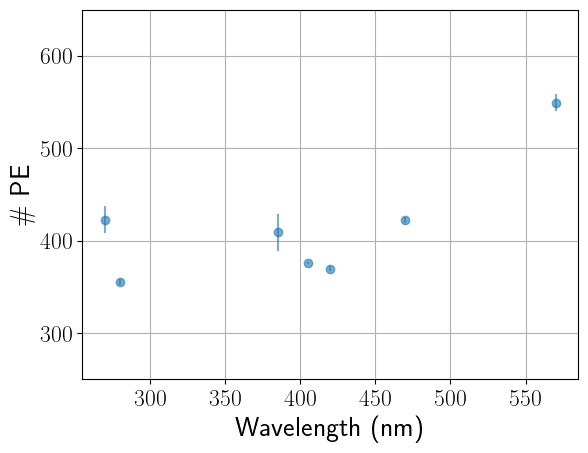}
    \caption{Mean number of detected PEs of the four individual measurements for light sources of different wavelengths. All the points are represented with error bars that represent the deviations observed in the measurements.}
    \label{fig:stability}
\end{figure}

For the proposed measurement a very stable light source is needed. We have performed dedicated measurements to ensure the repeatability of the amount of light sent to the SiPM. 
A reference sensor has been illuminated with light sources of different wavelengths in four separated data taking periods. The mean value of the detected light in the four measurements for each wavelength is shown in  Figure~\ref{fig:stability}. We can confirm the repeatability of the light intensity as the deviations are smaller than 1\% for all light sources except those of 270 and 385~nm, which are within 5\%. These errors have been propagated to the PDE calculation.

\subsubsection{Vacuum heat exchanger setup}
\begin{figure}[H]
    \centering
    \begin{subfigure}[b]{0.45\textwidth}
        \centering
        \includegraphics[width=\linewidth]{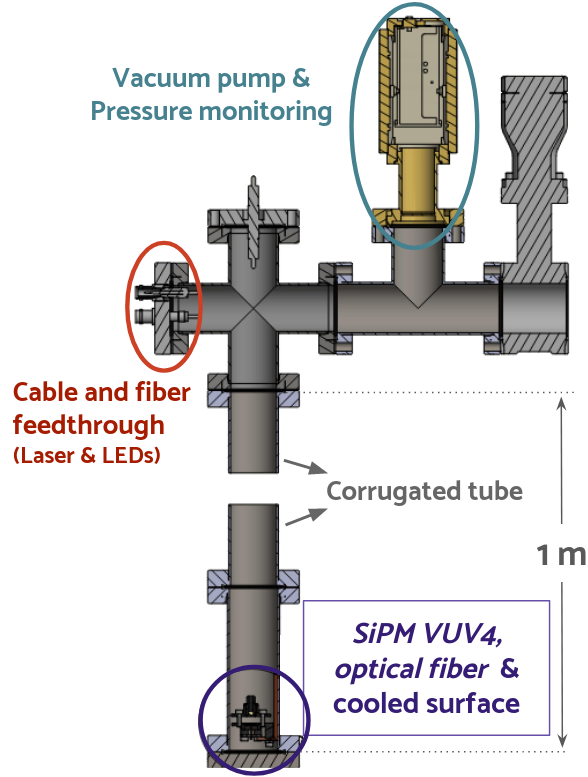}
        \caption{}
        \label{fig:vacuum_plano}
    \end{subfigure}%
    \begin{subfigure}[b]{0.37\textwidth}
        \centering
        \includegraphics[width=\linewidth]{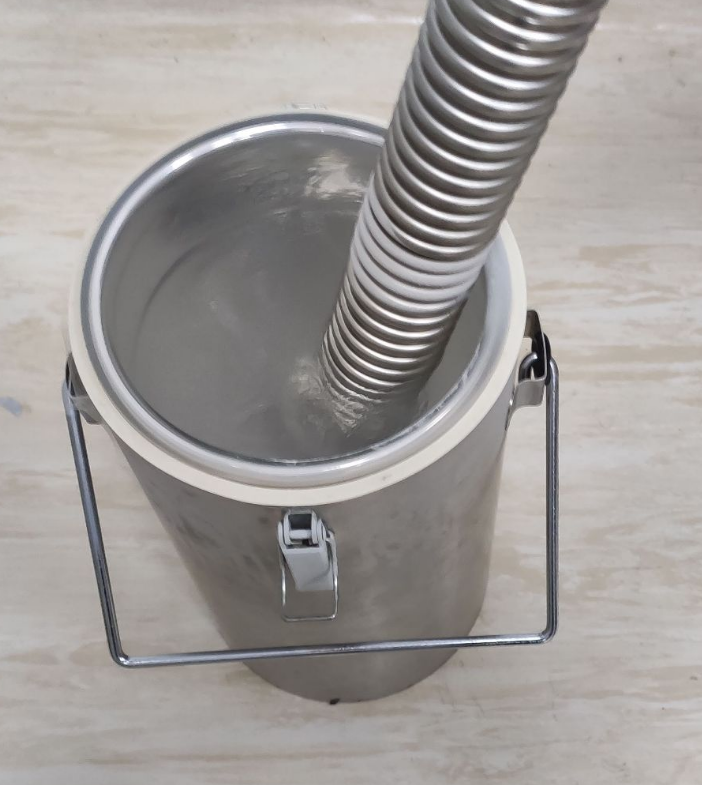}
        \caption{}
        \label{fig:coldfinger_photo}
    \end{subfigure}
    \caption{(a) Technical scheme of the mechanical structure of the vacuum cold finger setup. The vacuum is created in the whole stainless steel tube. The corrugated part is placed in the scheme gap of the rigid tube to join both parts in a flexible way. The heat exchanger is in the lower part of the tube, where the SiPM is allocated, and it is eventually submerged in LN$_2$. (b) The lower part of the tube with the SiPM inside a vessel filled with LN$_2$ to cool down the SiPM.}
\label{fig:COLD_FINGER_TECNICAL}
\end{figure}

The main goal of the setup is to provide a controlled mechanism to change the temperature of the photosensor without disrupting or altering the light propagation medium to ensure that the same amount of light reaches the SiPM independently of its temperature. The SiPM is cooled down with a heat exchanger, the so--called {\it cold--finger}; to thermally insulate the SiPM attached to the cold--finger, both the SiPM and the light source are placed in a tube at vacuum. The technical design scheme of this setup can be seen in Figure~\ref{fig:vacuum_plano}. The SiPM is placed at the end of a 1 m tube, where a $10^{-4}$~mbar vacuum is created. The power supply and the output analog signal cables are connected to the feed--through. To cool down the stainless steel tube, and thus the SiPM, its lower part is introduced into a small vessel filled with LN$_2$ as shown in Figure~\ref{fig:coldfinger_photo}.

\begin{figure}[H]
    \centering
    \includegraphics[width=\linewidth]{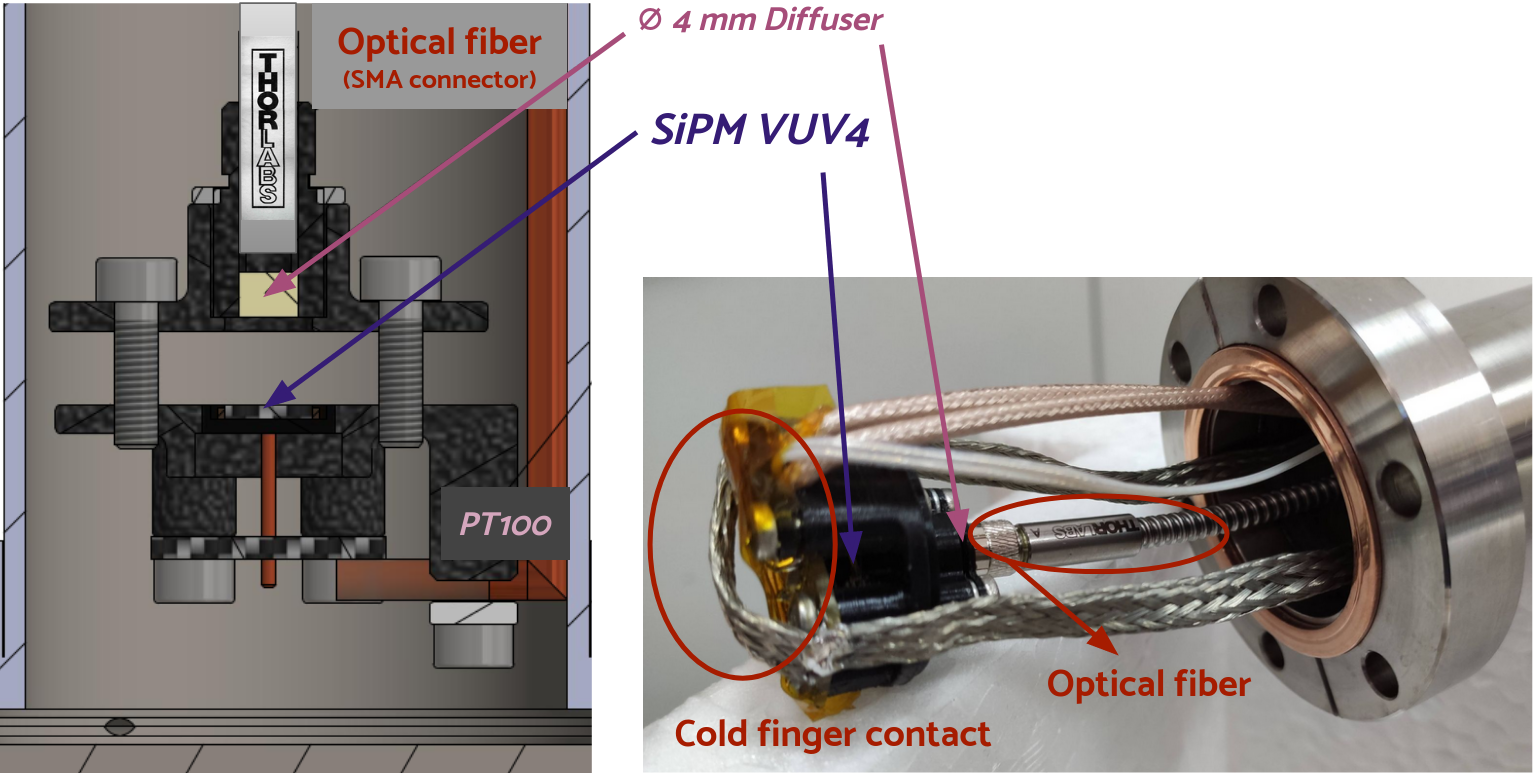}
    \caption{Left: Technical scheme of the mechanical structure of the vacuum cold finger setup. The diffuser, the optical fiber, the SiPM and the temperature sensor (PT100) are placed inside the tube. Right: Close--up view of the mechanical implementation of the vacuum pipe containing the SiPM photosensor and cold finger arrangement.}
    \label{fig:plano_vacuum}
\end{figure}

The SiPM is facing an optical fiber, which runs from the feed--through and is externally connected to the different light sources. The temperature sensor is a PT100 class A ($\pm 0.15 ^{\circ}C$ tolerance)  with a validity range that reaches 70 K. It is located in the structure that supports the SiPM as can be seen in Figure~\ref{fig:plano_vacuum} (Left). The SiPM temperature is considered to be the one indicated by the temperature sensor. The heat exchanger consists of a shielding wire mesh wrapped around the SiPM holding bracket in contact with the tube (see Figure~\ref{fig:plano_vacuum} (Right)). This part is submerged (once closed) in LN$_2$ to cool down the SiPM by thermal contact with the tube.

\subsubsection{GAr vessel setup}
A second setup has been designed to cross--check the results. In this case, we have improved the temperature control and include the feasibility for operating with GAr together with an $^{241}$Am source. This will allow us to determine the PDE at 127~nm in forthcoming works. The determination of produced scintillation light from excited gaseous argon is very challenging because of its dependence on temperature and geometrical factors and requires further development. However, the setup can also be used for the measurement of the PDE for visible light. 
This GAr setup enables the possibility of having an homogeneous and stable environment at both room and cryogenic temperatures and water condensation is avoided without the need of making vacuum in the setup.


\begin{figure}[H]
    \centering
    \begin{subfigure}[b]{0.45\textwidth}
         \centering
    \includegraphics[width=\linewidth]{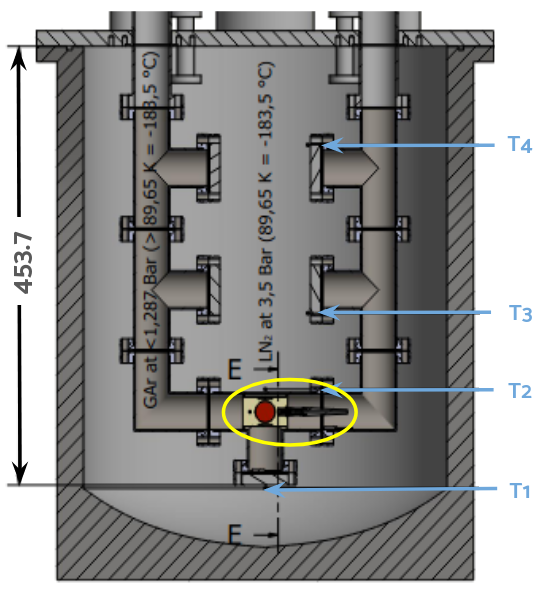}
    \caption{}
    \label{fig:gar_plano}
    \end{subfigure}%
    \begin{subfigure}[b]{0.45\textwidth}
        \centering
    \includegraphics[width=\linewidth]{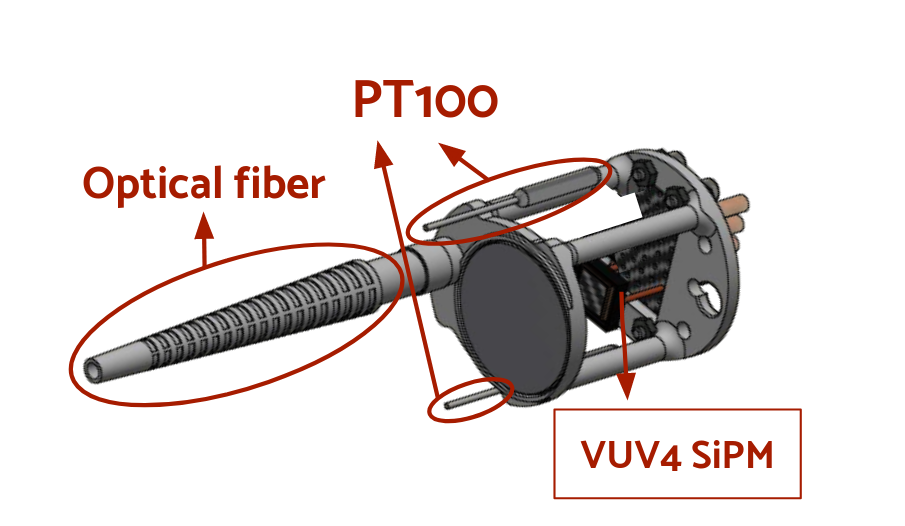}
    \caption{}
    \label{fig:gar_zoom}
    \end{subfigure}
    \caption{(a) Technical design of the GAr setup with dimensional information in mm. (b) Zoom on the structure (yellow circle on the left) where the SiPM, the PT100 sensors and the optical fiber are installed.}
    \label{fig:GAr_VESSEL}
\end{figure}

A schematic representation of the setup can be seen in Figure~\ref{fig:gar_plano}. It consists of a 50~l vessel that holds a U--shaped tube through which GAr is continuously circulating. In the center of the tube there is a holder where the optical fiber is fixed facing the SiPM (see Figure~\ref{fig:gar_zoom}). A light diffuser is used to homogenize the light reaching the SiPM from the fiber, reducing the impact of the different refraction index for warm and cold GAr. Due to the short distance between the fiber and the SiPM, we can assume that the same amount of light is reaching the SiPM at both RT and CT.

Once the setup is assembled, vacuum is created in the tube where the photodetector was located, reaching levels of $10^{-4}$~mbar. Then the GAr is allowed to circulate continuously through the tube. The gas is expelled to the outside medium through a non--return valve that prevents contamination with outside air. After taking data with several wavelengths light sources at RT, the 50~l vessel is filled with LN$_2$ to cool the U--tube, and therefore the SiPM by thermal contact. Four thermal sensors located at different heights were used to monitor the temperature change throughout the vessel. The required measurements were successfully repeated at different cryogenic temperatures and the results are described in section~\ref{sec:results}.  \medskip

\subsection{Setup for PDE measurement for 127 nm light}

\begin{figure}[H]
    \centering
    \begin{subfigure}[b]{0.38\textwidth}
    \includegraphics[width=\linewidth]{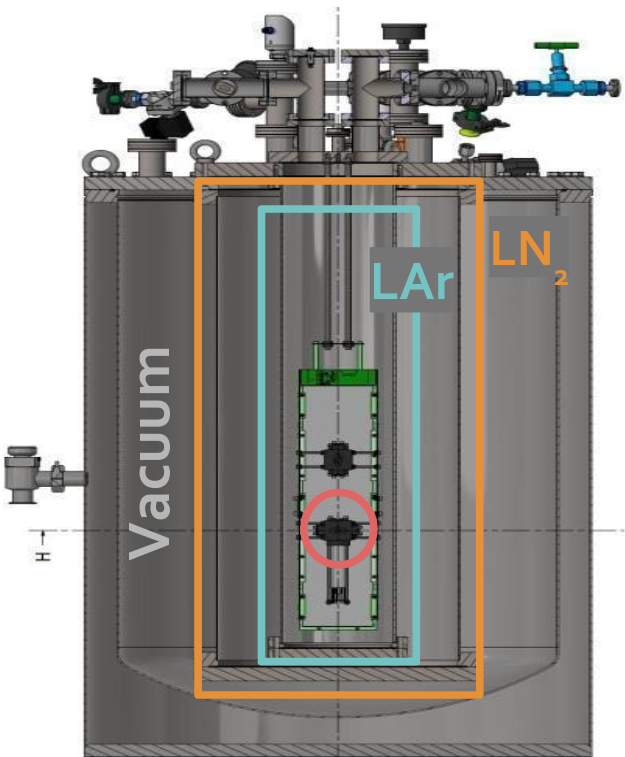}
    \caption{}
    \label{fig:setup_LAr_a}
    \end{subfigure}%
    \begin{subfigure}[b]{0.62\textwidth}
    \includegraphics[width=\linewidth]{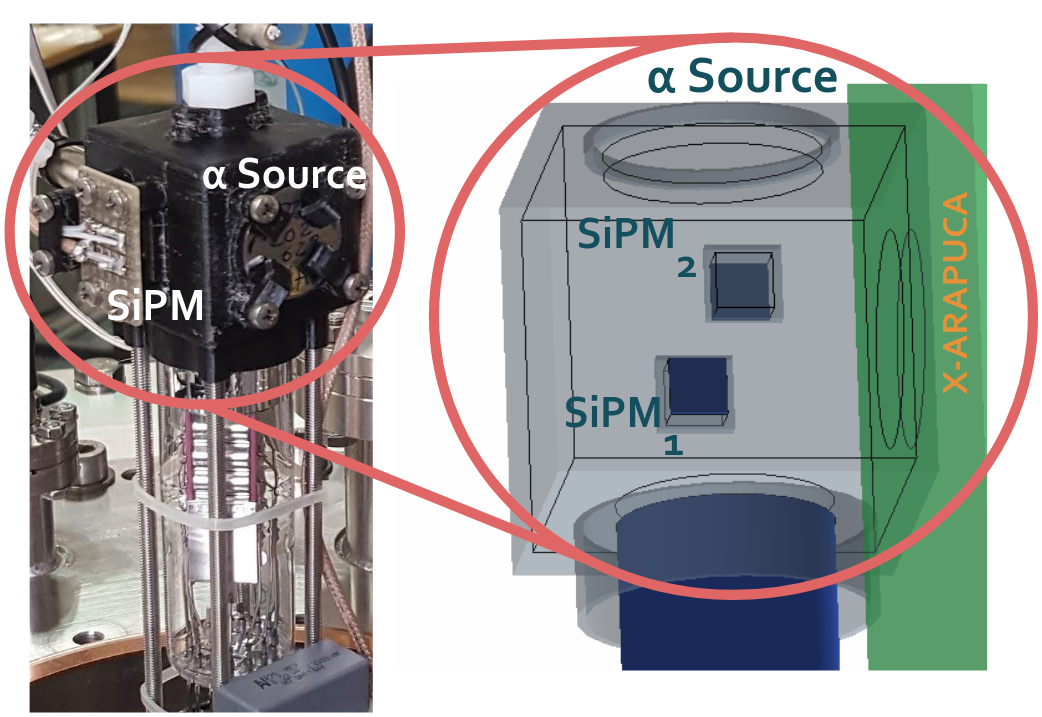}
    \caption{}
    \label{fig:setup_LAr_b}
    \end{subfigure}
    \caption{Setup scheme used for obtaining the absolute efficiency of the VUV4 SiPMs at 127~nm. (a) Cryogenic vessel with its concentric volumes where GAr is liquified. (b) Picture of the black box holding the sensors together with the $^{241}$Am source with its corresponding diagram.}
    \label{fig:setup_LAr}
\end{figure}

This measurement is done in the cryogenic setup prepared for the PDE measurement of the DUNE X--ARAPUCA~\cite{2104.07548_X-ARAPUCA_PD_DUNE}.
In this setup, we made use of a 300~l cryogenic vessel with different concentric volumes, which schematic is shown in Figure~\ref{fig:setup_LAr_a}. There are two concentric internal volumes, a larger one (100~l), where the LN$_2$ is introduced and a smaller one (18~l), where the VUV4 SiPMs are located together with the X--ARAPUCA, and filled with GAr. In this 18~l container, the GAr is liquefied by thermal contact with the LN$_2$ of the surrounding volume. This is achieved by controlling the pressure parameters that regulate the temperature values necessary to carry out the liquefaction. 
The GAr 99.9999\% is liquefied with LN$_2$ at 2.7~bar. To avoid outgassing we perform successive vacuum cycles before introducing the optical and electrical components that will be used to perform the measurements.
The 127~nm scintillating light is produced by an $^{241}$Am alpha source placed together with two SiPMs in an opaque box, as shown in Figure~\ref{fig:setup_LAr_b}. 
It emits $\alpha$ particles with 5.485~MeV (84.45\%) and 5.443~MeV (13.23\%) energies with an activity of (54.53 $\pm$ 0.82)~Bq~\cite{2023.110913_Validation241AmLiquifiedGasDetectorsCT}. The $\alpha$ particles deposit their energy inside the 4~cm sized black box, and the produced photons reach the VUV4 SiPMs. The ADC acquisition is triggered by the coincidence of the two SiPMs signal above a certain threshold. The setup includes an optical fiber attached to a LED that allows the calibration of the SiPMs in the single PE regime. In this case, the data taking is triggered by the pulse generator synchronously with the LED. For this setup the SiPMs have been biased at 4~V over the breakdown voltage measured in LN$_2$ corresponding to a 3.5~OV at LAr temperature. In order to compare with the manufacturer data a correction factor of 3\% should be applied.

\section{Results} \label{sec:results}

In the following section, the experimental results obtained with the setups introduced in section~\ref{sec:setup} are presented. The PDE for visible light is determined from the ratio of the light collected at CT and RT at a light regime for which we have demonstrated a constant illumination of the sensors. Since the gain and other operation parameters of the SiPMs depend on the temperature, the estimation of the amount of PEs detected is crucial for a reliable measurement of the PDE. First of all, the SiPM gain at CT and RT is computed from the calibration runs. Then a dedicated study on the cross--talk at CT is presented and finally, the results for the PDE are discussed.

\subsection{Gain calibration}\label{sec:calibration}

The gain calculation relies on the calibration of the sensor at the single photo--electron level. While at CT we have a well--determined baseline and a clear signal, at RT there is much more dark--count noise making the integration of individual pulses more complicated. Therefore the data sample is filtered to get events with a single pulse and a clean baseline in the integration window. 

\begin{figure}[H]
    \centering
    \includegraphics[width=0.75\linewidth]{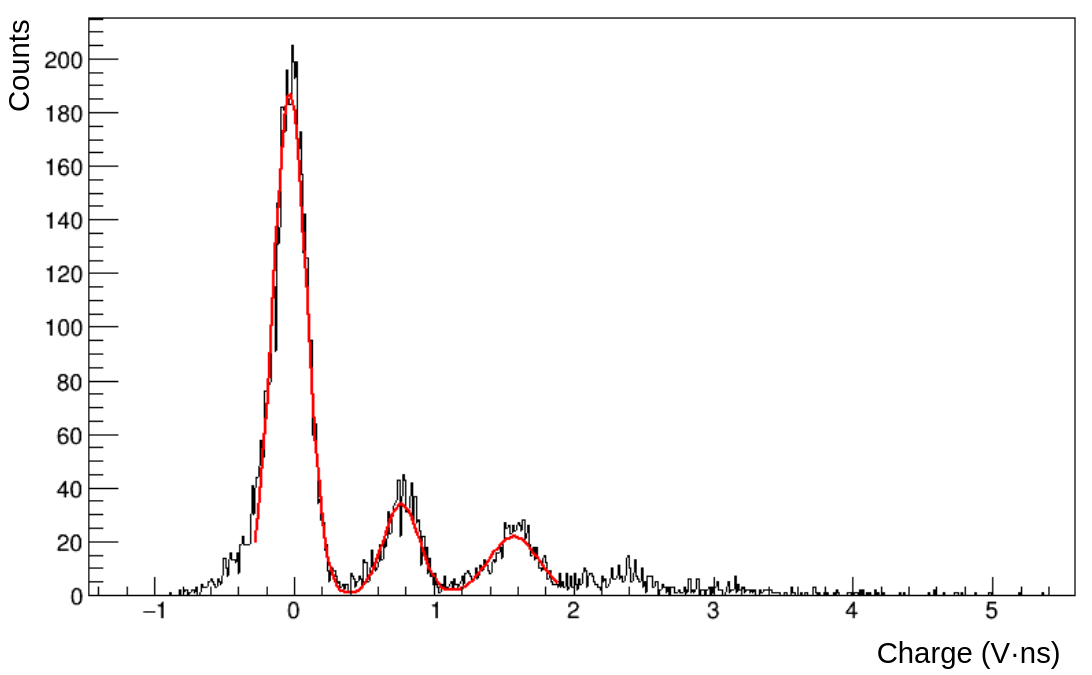}
    \caption{Example charge spectrum corresponding to a calibration at OV $= 3$ V at RT with a three Gaussian fit.}
    \label{fig:CHARGE_SPECTRUM}
\end{figure}

Experimentally, the gain can be obtained from the charge spectra after integrating the individual pulses.
These spectra are fitted to N Gaussians, corresponding to the electronic noise (pedestal) and the charge deposited by 1... N-1 PE. This is presented in Figure~\ref{fig:CHARGE_SPECTRUM} for the lowest OV and at RT.

The distance between consecutive maximums of the Gaussians is the gain of the system.
The error accounts for the dispersion of the measurements for different calibration runs. 

Mean gain values are obtained at different temperatures for each OV and, as expected, the gain at CT is higher than at RT.


\subsection{Correlated noise including cross--talk and afterpulses} \label{sec:x-talk}

The number of PEs detected by the SiPM should be corrected by the cross--talk probability (that is the probability that an avalanching microcell causes an avalanche in a second microcell). It is known that the cross--talk probability changes with temperature~\cite{1903.03663_VUV4S13370nEXO,1981_Dumke_SiPerformanceCT}. So, it is critical to correctly account for this effect in the PDE determination. 

\begin{figure}[H]
    \centering
    \includegraphics[width=0.65\textwidth]{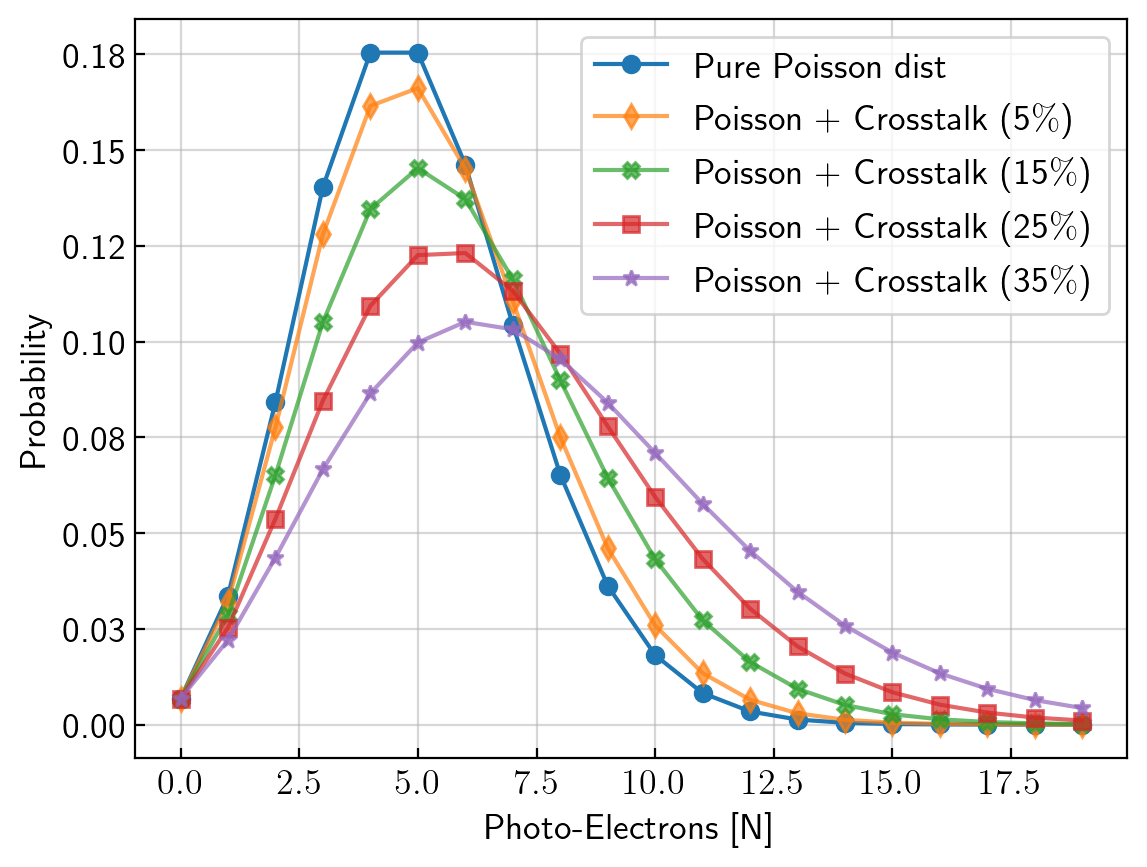}
    \caption{Compound Poisson distribution function for $p=$ [0.05, 0.15, 0.25, 0.35] being $p=0$ the pure Poisson distribution.}
    \label{fig:PurePoissonian_comparisson}
\end{figure}

We have performed a dedicated analysis to extract the cross--talk value for the VUV4 SiPMs (series: S13370) at CT. To describe the joint effect of cross--talk and afterpulses in the PE distribution we have made use of a composite Poissonian as described in~\cite{N25-111_NSS-MIC_2009_NoiseFactor-CT+afterpulses_Vinogradov,1109.2014_Vinogradov}. 
Figure~\ref{fig:PurePoissonian_comparisson} represents the different predicted distributions for the normalized charge histograms depending on the cross--talk value and how it compares with a pure Poisson distribution.

\begin{figure}[H]
\centering
    \begin{subfigure}{0.5\linewidth}
        \centering
        \includegraphics[width=\linewidth]{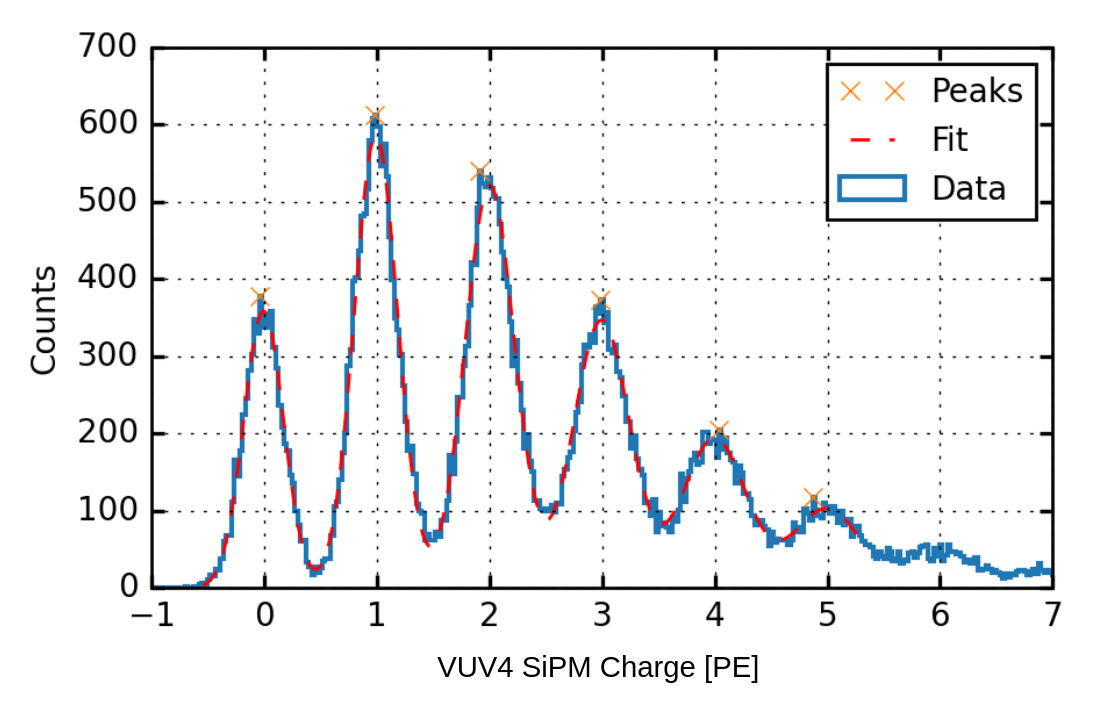}
        \caption{}
        \label{fig:finger_plot_2}
    \end{subfigure}%
    \begin{subfigure}{0.5\linewidth}
        \centering
        \includegraphics[width=\linewidth]{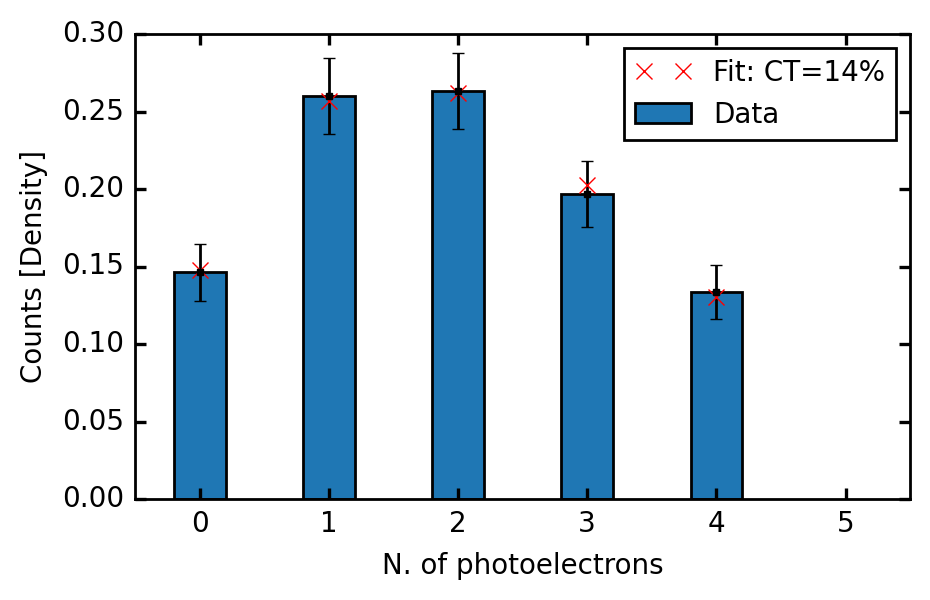}
        \caption{}
        \label{fig:vinogradov_cross-talk}
    \end{subfigure}
    \caption{(a) Charge histogram for a calibration run at OV $=4$ V. (b) Normalized density charge histogram fitted with a Vinogradov fit to extract the cross--talk value at CT.}
    \label{fig:cross-talk}
\end{figure}

The cross--talk probability~($P_{CT}$) at CT is defined as the number of events above 1.5 PE divided by the total number of events. It is computed with a calibration run with enough light to obtain the plot in Figure~\ref{fig:cross-talk} and perform a Vinogradov fit to the mentioned composite Poissonian. The correction factor for the PDE considering the cross--talk contribution of higher order is computed following the model described in~\cite{N25-111_NSS-MIC_2009_NoiseFactor-CT+afterpulses_Vinogradov}. It can be estimated as $f_{CT} = PE_{det}/PE$ where $PE_{det}$ is the number of detected photons and $PE$ is the number of photons that would be detected in the absence of cross--talk. By considering the chain of secondary pulses produced by a single primary pulse, the $PE_{det}$ can be expressed as $PE$ multiplied by a geometric distribution with mean $P_{CT}$. Therefore, the correction factor is $f_{CT} = (1-P_{CT})$.


The $P_{CT}$ computed for this SiPM at CT and an OV 4V is $(14.84 \pm 0.24)\%$ which is in perfect agreement with the presented value in~\cite{2202.02977_SiPMVUV4_PDE_CT+128nm}. 
At RT we use the value measured by HPK of $5\%$
because after filtering the signal, the resulting charge histogram does not have enough statistics for performing a Vinogradov fit.

\subsection{Cryogenic Photon Detection Efficiency computation} \label{sec:PDE_calculation}

\subsubsection{PDE measurement for visible light}

The amount of light detected by the SiPM at a given overvoltage is measured in terms of PEs, and is determined as the integrated charge in a 3~$\mu$s waveform divided by the SiPM gain at the corresponding temperature for this overvoltage. 

The ratio of the number of PEs measured at RT and CT for the same light intensity at a given wavelength is used to determine the PDE at CT as shown in
equation~\ref{eq:PDE_CT}. The obtained value should be corrected by the increase of the cross--talk at CT. 

\begin{figure}[H]
    \centering
    \includegraphics[width=\textwidth]{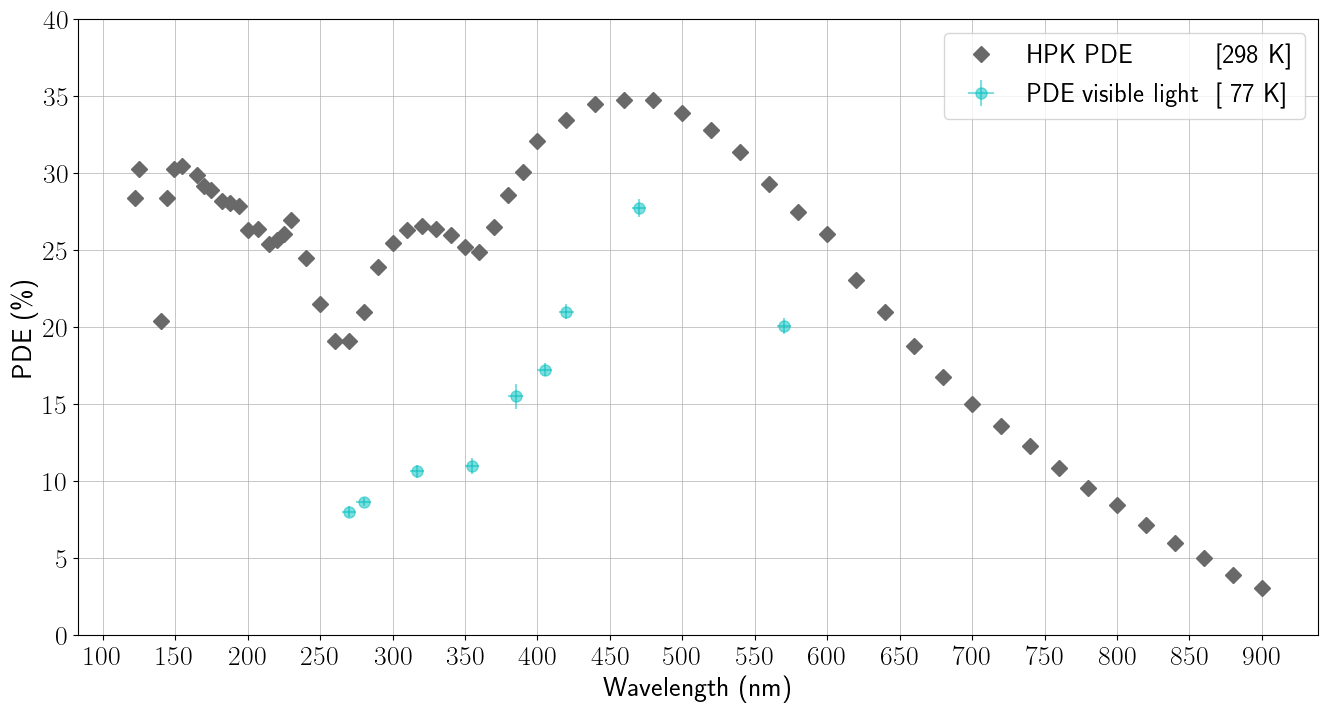}
    \caption{PDEs values at RT (grey diamonds) and at CT  for vacuum setup (green circles).}
    \label{fig:relative_pde_1}
\end{figure}

The PDE values obtained in the visible wavelength range are tabulated in Table~\ref{tab:pde_ct} and shown in Figure~\ref{fig:relative_pde_1} for one SiPM. 

\begin{table}[!ht]
\centering
\begin{tabular}{cccc} 
 \hline
    $\lambda$ (nm) & PDE$_\text{RT}$ ($\%$) & PDE$_\text{CT}$ ($\%$) & Ratio\\  \hline  \hline
        270 & 19.10 & 8.05 \ ± 0.36 & 0.42 ± 0.02 \\
        280 & 21.00 & 8.66 \ ± 0.25 & 0.41 ± 0.01 \\
        317 & 26.45 & 10.67 ± 0.41 & 0.40 ± 0.02 \\
        355 & 25.05 & 11.00 ± 0.49 & 0.44 ± 0.02 \\
        385 & 29.35 & 15.52 ± 0.83 & 0.53 ± 0.03 \\
        405 & 32.10 & 17.27 ± 0.36 & 0.54 ± 0.01 \\
        420 & 33.50 & 21.02 ± 0.44 & 0.63 ± 0.01 \\
        470 & 34.80 & 27.76 ± 0.53 & 0.80 ± 0.02 \\
        570 & 28.40 & 20.11 ± 0.51 & 0.71 ± 0.02 \\ \hline
    \end{tabular}
    \caption{PDE for both temperature regimes. The RT one is given by the manufacturer, while the CT one is the obtained after the analysis described in this document.}
    \label{tab:pde_ct}
\end{table}

\begin{figure}[H]
    \centering
    \includegraphics[width=0.65\textwidth]{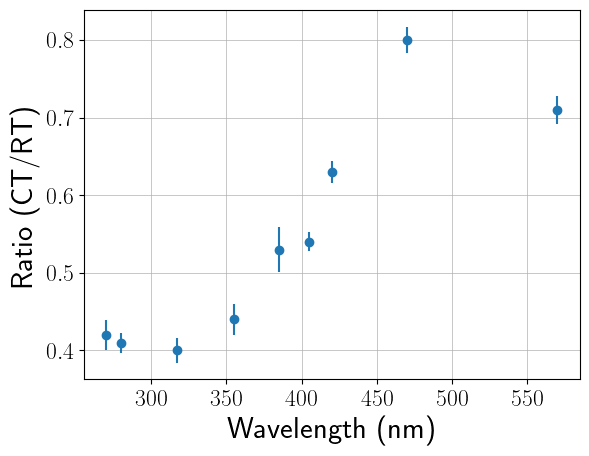}
    \caption{Ratio between the PE measured at CT and RT for the studied range of wavelengths.}
    \label{fig:ratio}
\end{figure}

We can see that the PDE at CT shows a clear decrease compared to its value at RT. Not only that, but there is a dependence with the wavelength that is presented in Figure~\ref{fig:ratio}.

We can clearly notice a minimum at $\sim$300~nm and a maximum at $\sim$500~nm.

The same procedure is followed for the GAr setup, where in order to ensure that the light reaching the SiPMs is independent of the temperature of the setup, we allocate the optical fiber close enough to the sensors and  additionally we use a diffuser so that the light is emitted homogeneously. 

\begin{figure}[H]
    \centering
    \includegraphics[width=\textwidth]{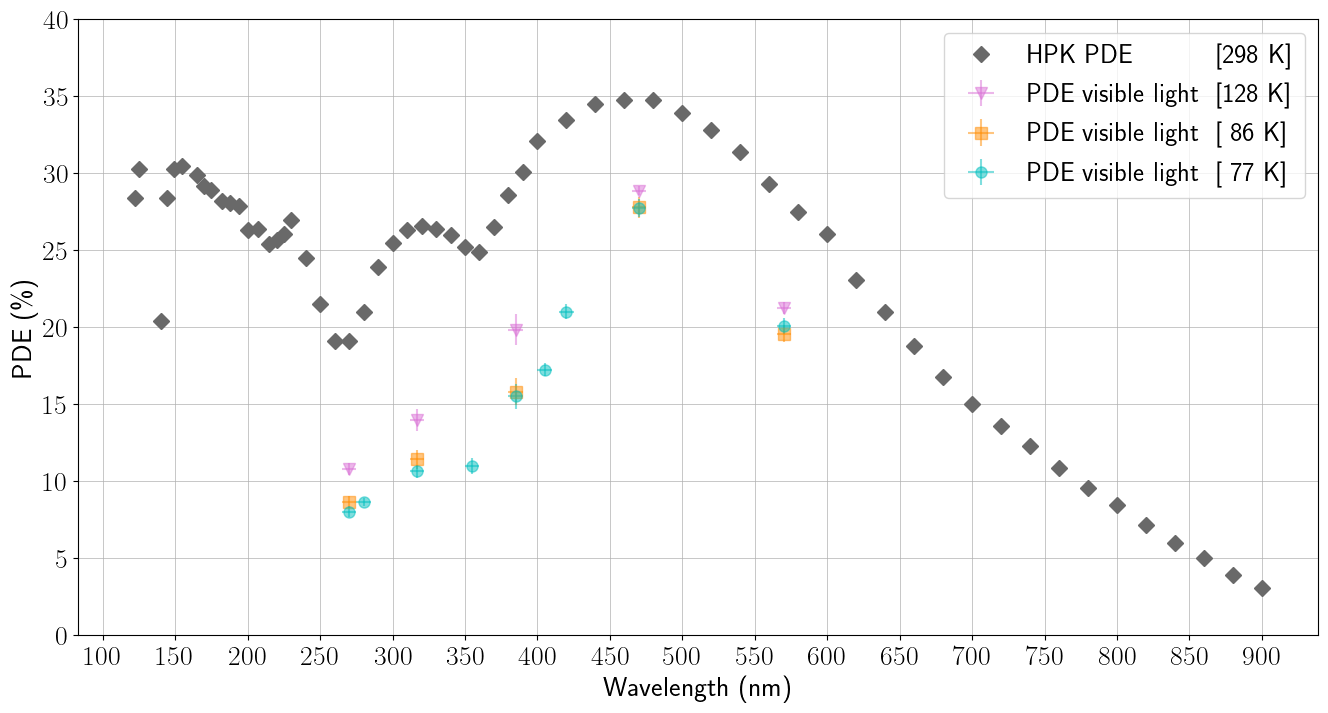}
    \caption{PDEs values at RT (grey diamonds), for GAr setup at 128~K (pink down triangles) and at 86~K (orange squares) and for vacuum setup at 77~K (green circles).}
    \label{fig:relative_pde_intermediate}
\end{figure}

Additionally, as we have increased the number of temperature sensors from one to four, we have a better determination of the setup temperature. 
This allows us to do measurements at intermediate temperatures during cooling.
In Figure~\ref{fig:relative_pde_intermediate} we can see the results obtained in GAr for two cryogenic temperatures,  86~K (equilibrium) and (128$\pm$2)~K (during thermalization).

We can clearly notice the same decrease of PDE with temperature, validating our results at 77~K. The measurement at 128~K gives additional information on the temperature--dependence of the PDE.

\subsubsection{PDE measurement for 127~nm light}

\begin{figure}[htp]
    \centering
    \includegraphics[width=0.5\textwidth]{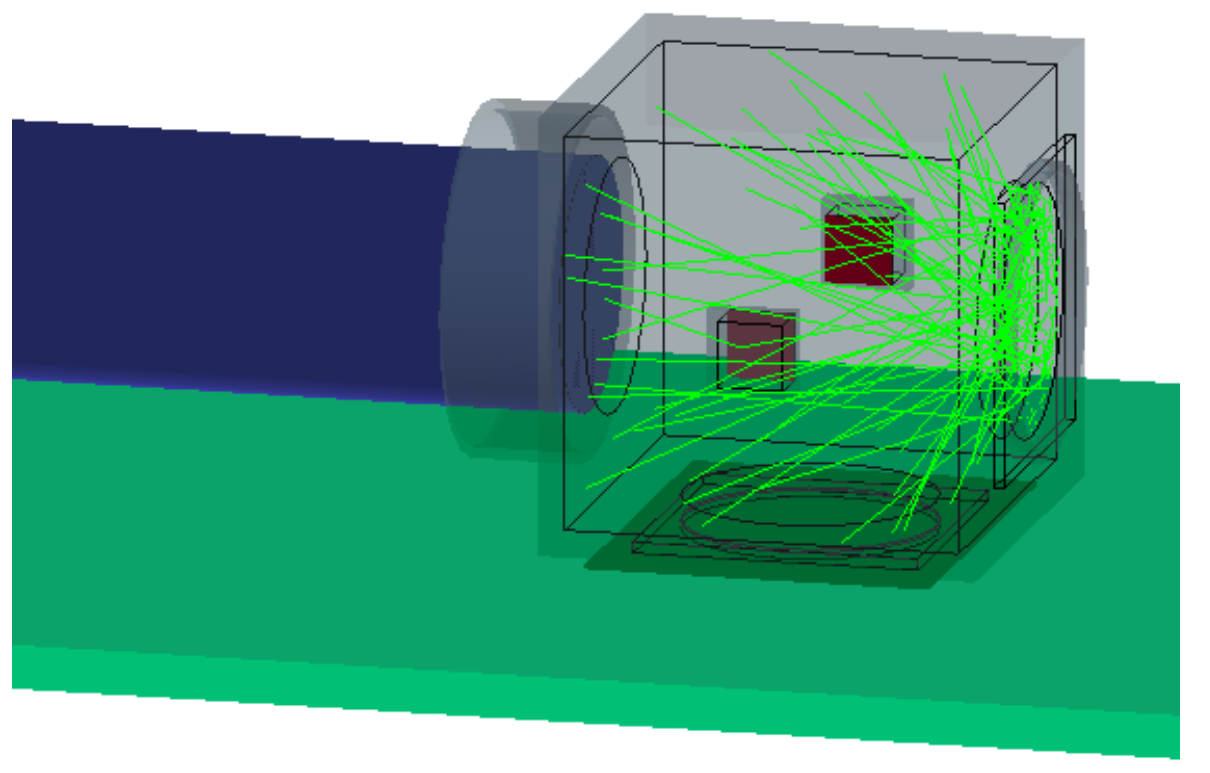}
    \caption{Simulation of optical photons from an extense source to study their geometrical distribution and the number reaching the VUV4 SiPMs with GEANT4 \cite{GEANT4}.}
    \label{fig:simulation}
\end{figure}

As introduced in section~\ref{sec:methodology}, we have performed another measurement by submerging the SiPMs in LAr together with an $^{241}$Am source to measure the PDE at CT and at 127~nm. For this computation, we compare the light collected by the SiPM with the expected number of PEs taking into account the position of the SiPMs relative to the source and the scintillation light--yield of an $\alpha$ particle of 5.48~MeV in LAr.
To determine the fraction of the produced scintillation light that hits the SiPM area, a detailed simulation of the 4~cm side black box, including the dimensions of the source ($\oslash = 20$~mm) and a realistic simulation of the sensor holders, is made. Figure~\ref{fig:simulation} features a view of the simulated elements and the trajectories of the isotropic generated photons.

\begin{figure}[htp]
    \centering
    \includegraphics[height=0.55\linewidth]{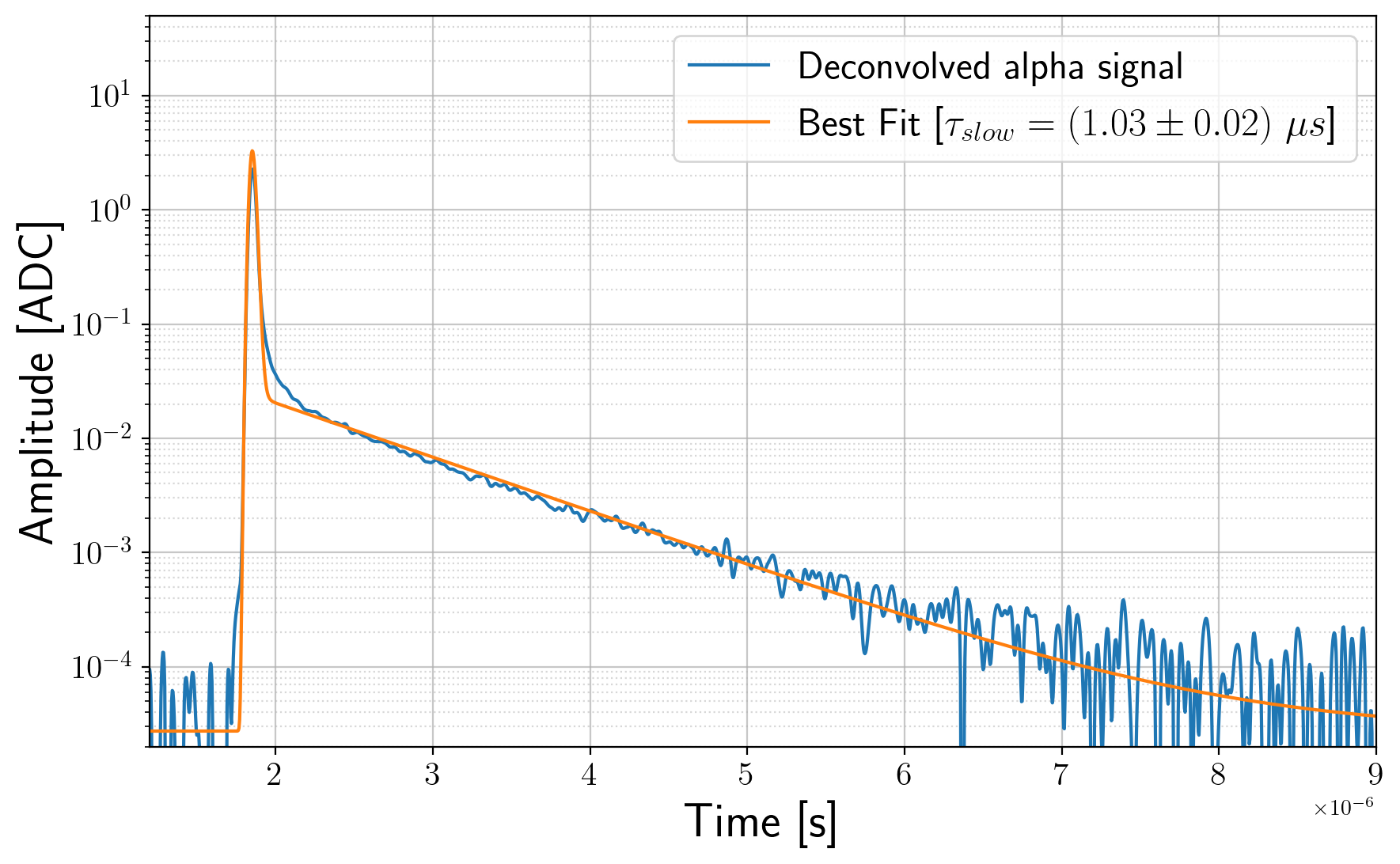}
    \caption{Fit of the deconvoluted scintillation profile for the VUV4 SiPM.}
    \label{fig:purity_fit}
\end{figure}

\vspace{0.6cm}
The number of generated scintillation photons is determined considering a light--yield of 51000~photons/MeV~\cite{Doke1981_FundamentalProperties} and a quenching factor of 0.70 for $\alpha$ particles~\cite{CarvalhoKlein_LuminiscenseCondesedArgon}. 
The increase of the cross--talk at CT is corrected as in the measurement for visible light. Finally, the quenching of the scintillation light production by impurities in the LAr that affects the triplet emission with a longer decay time is computed by following the next equation~\cite{Acciarri_2009_N2+O2inLAr}:
\begin{equation}
    f^{-1}_{\text{purity}} \equiv A_{\text{triplet}} \frac{\tau^{\text{exp}}_\text{{slow}}}{\tau^{\text{th}}_\text{{slow}}} + A_{\text{singlet}} = (0.94 \pm 0.02) \quad ,
\end{equation}

where $\rm \tau^{th}_{\text{slow}}=1.41\ \mu$s, $\rm A_{triplet}=0.23$, $\rm A_{singlet}=0.77$ are the theoretical parameters describing an ideal scintillation profile and are obtained from the literature \cite{2104.07548_X-ARAPUCA_PD_DUNE,Hitachi1983}. The measured $\tau^{\text{exp}}_{\text{slow}}$ is obtained by fitting the deconvoluted scintillation profile of the SiPMs as shown in Figure~\ref{fig:purity_fit}.

As introduced in section~\ref{sec:setup}, we have made use of two VUV4 SiPMs that have been calibrated as in section~\ref{sec:calibration}. For obtaining the charge associated to the LAr scintillation photons we acquire signals from a coincident trigger, resulting in an averaged charge of $(55.00 \pm 1.99)$~PE. The expected number of photons in each SiPM after applying the purity factor is $495$~PE. With the cross--talk and OV corrections, the estimated PDE at CT for the LAr scintillation light is:
\begin{equation*}
    \text{PDE}_{\text{127 nm}} = (12.69 \pm 1.12) \% \quad .
\end{equation*}
The PDE has been computed individually for the two SiPMs and the shown result is the mean value. We can see that the absolute measurement for 127~nm shows a decrease in PDE at CT. Moreover, this value can be compared with the recently published value in~\cite{2202.02977_SiPMVUV4_PDE_CT+128nm}, $(14.7^{+1.1}_{-2.4})$\%. 
In this study they have subtracted the reflected photons ($24\%$) from the contribution of photons hitting the sensor, contrary to our method.
The result is compatible within errors with our current result if the same definition is applied. The main uncertainty of our measurement is $8.3\%$ coming from the geometrical tolerance of our setup, i.e., the uncertainty on its position, followed by a $1.9\%$ uncertainty due to the purity correction.

\section{Conclusions} \label{sec:conclusions}

\begin{figure}[H]
    \centering
    \includegraphics[width=\textwidth]{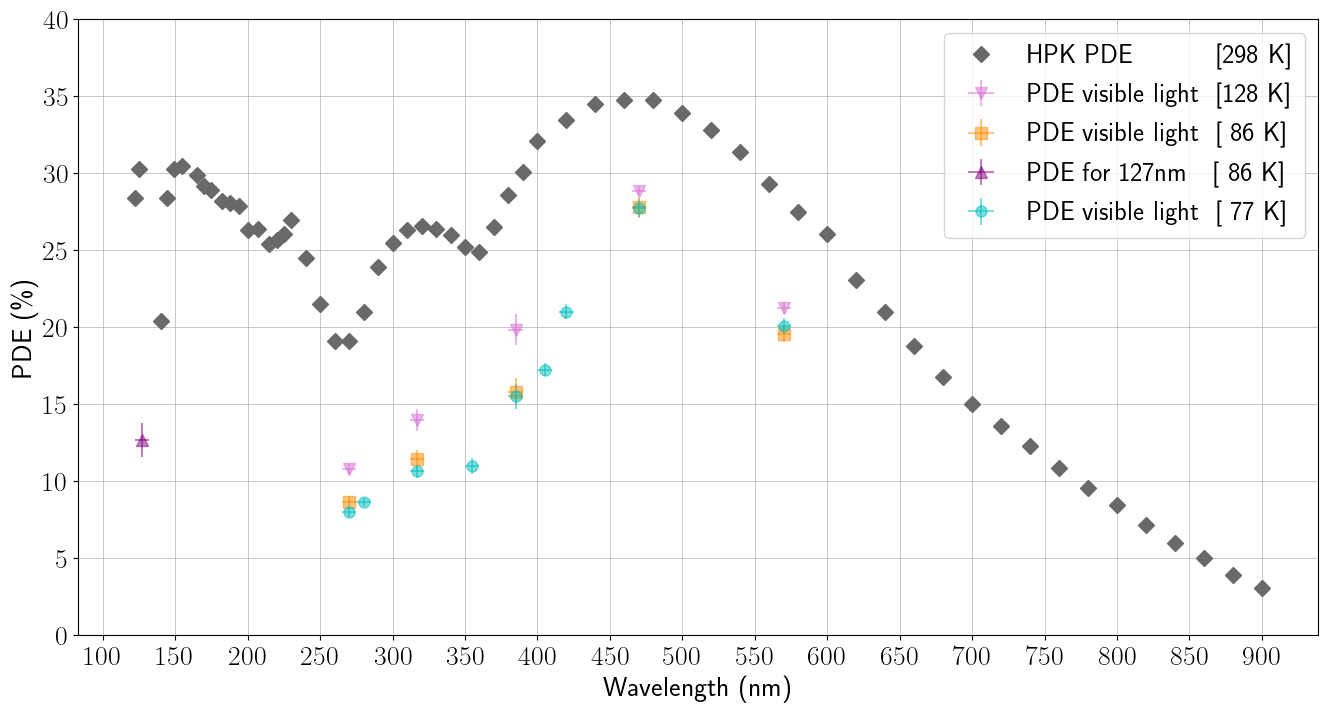}
    \caption{PDEs values at RT (grey diamonds), at 128~K for GAr setup (pink down triangles), at 86~K for GAr setup (orange squares) and at 86~K for LAr setup (purple up triangle) and at 77~K for vacuum setup (green circles).}
    \label{fig:relative_pde_3}
\end{figure}

In this work, we have presented the measurement of the absolute PDE of HPK VUV4 S13370--6075CN SiPMs at CT for different wavelengths in the range [270, 570]~nm and at 127~nm with a detailed description of the setups developed for this measurement. The results are summarized in Figure \ref{fig:relative_pde_3}.

We can clearly notice that for the three presented setups we obtain compatible results. The most remarkable conclusion is the decrease of the PDE for these SiPMs when operating at CT.
The PDE reduction at CT depends on the wavelength, being maximal ($\sim$60\%) at lower wavelengths ($<$350~nm) and about 20\% for wavelengths around 450 nm. The PDE has been measured at three different cryogenic temperatures: 77~K, 86 K and 128~K. The sensitivity of the measurements allows to observe a higher PDE at 128~K.
Besides that, this study confirms the value of the PDE at 127~nm presented in previous works.

\section*{Acknowledgments} \label{sec:Acknowledgments}

The present research has been supported and partially funded by European Union NextGenerationEU/PRTR, European Union’s Horizon 2020 Research and Innovation programme under Grant Agreement No. 101004761 and under the Marie Sklodowska-Curie grant agreement No 892933, by MCIN/AEI/10.13039/501100011033 under Grants No. PID2019--104676GB--C31, RYC2021--031667--I, PRE2019--090468 and PRE2020--094863 of Spain. Work produced with the support of a 2023 Leonardo Grant for Researchers in Physics, BBVA Foundation.

\clearpage
\bibliographystyle{elsarticle-num} 
\bibliography{references}
\end{document}